\documentclass[article,singlecolumn,superscriptaddress,preprintnumbers,nopacs,floatfix,amsmath,amssymb]{revtex4}

\usepackage[english]{babel}	
\makeatletter\AtBeginDocument{\let\@elt\relax}\makeatother 
\usepackage{amsmath}		
\usepackage{amssymb}		
\usepackage{bbm}			
\usepackage{graphicx}		
\usepackage{graphics}		
\usepackage[mathscr]{euscript}
\usepackage{calligra}

\newcommand{\la}[1]{\mbox{$
\lefteqn{ \mbox{\,\, \tiny #1}}$} \label{#1}}

\newcommand{\be}{\begin{eqnarray}}
\newcommand{\ee}{\end{eqnarray}}
\newcommand{\br}{\begin{matrix}}
\newcommand{\fr}{\frac}
\newcommand{\ds}{\begin{array}{llclllllll}}
\newcommand{\dst}{\end{array}}
\newcommand{\LRa}{\Leftrightarrow}

\newcommand{\er}{\end{matrix}}

\newcommand{\da}{\left(\begin{array}{llcllllllll}}
\newcommand{\dt}{\end{array}\right)}

\newcommand{\acc}{\\[3mm]}

\newcommand{\Vx}{{\mathbf x}}

\newcommand{\eastar}{\end{eqnarray*}}

\begin{document}

\title{The SU(2) Lie-Poisson Algebra and its Descendants}

\author{Jin Dai}
\email{Jin.Dai@su.se}
\affiliation{Nordita, Stockholm University and Uppsala University, Hannes Alfv\'ens v\"ag 12, SE-106 91 Stockholm, Sweden}
\affiliation{Department of Physics, Beijing Institute of Technology, Haidian District, Beijing 100081, P. R. China}
\author{Theodora Ioannidou}
\email{ti3@auth.gr}
\affiliation{
Faculty of Civil Engineering,  School of Engineering, Aristotle University of Thessaloniki, 54249, Thessaloniki, Greece
}
\author{ Antti J. Niemi}
\email{Antti.Niemi@su.se}
\affiliation{Nordita, Stockholm University and Uppsala University, Hannes Alfv\'ens v\"ag 12, SE-106 91 Stockholm, Sweden}
\affiliation{Laboratoire de Mathematiques et Physique Theorique CNRS UMR 6083,
F\'ed\'eration Denis Poisson, Universit\'e de Tours, Parc de Grandmont, F37200, Tours, France}
\affiliation{Department of Physics, Beijing Institute of Technology, Haidian District, Beijing 100081, P. R. China}

\begin{abstract}
\noindent
In this paper, a novel discrete algebra is presented which follows by combining the SU(2) Lie-Poisson bracket with the discrete Frenet equation. 
Physically,   the construction  describes a discrete piecewise linear string in $\mathbb R^3$. 
The starting point of our derivation is  the  discrete Frenet frame assigned at each vertix of the string.
 Then  the link vector that connect the neighbouring vertices is assigned the SU(2)  Lie-Poisson bracket.
Moreover, the same bracket defines the  transfer matrices of the discrete Frenet equation which  relates two neighbouring frames along the string. 
The procedure extends in a self-similar manner  to  an  infinite hierarchy of  Poisson structures.  
As an example,  the first descendant of the SU(2) Lie-Poisson structure is presented in detail. 
For this, the spinor representation of the discrete Frenet equation is employed, as it converts the brackets into a computationally more manageable form. The final result  is a  nonlinear,  nontrivial and novel Poisson structure  that engages four neighbouring vertices.  

\end{abstract}

\pacs{}

\maketitle

\section{Introduction}

The Poisson structure \cite{Poisson-2013}
is a widely investigated concept that has both physical and mathematical relevance.  The concept originates from Poisson's 
research on analytic mechanics, which  now provides a very general and solid  framework for describing Hamiltonian dynamics.
Mathematically, a Poisson structure associates to  every smooth function $H$ on a smooth manifold $\mathcal M$, a vector field $\mathcal X_H$. 
This vector field determines Hamilton's equation of motion, while the function  $ H$ is the so-called  Hamiltonian. 
Whenever the  pertinent Poisson bracket is also a Lie bracket, it 
ensures the validity of Poisson's theorem that states that  the Poisson bracket of two constants of motion is itself a constant of motion.

The SU(2) Lie-Poisson bracket is a classic example of a Poisson bracket structure, originally introduced by Lie \cite{Lie-1874}. 
However,  its systematic investigations came much later, and  started with the seminal work by Lichnerowicz \cite{Lichnerowicz-1977} who also introduced the  concept of a Poisson structure. 
Important early contributions to the development of Poisson structures were made by 
 Kirillov \cite{Kirillov-1976} and in particular by  Weinstein \cite{Weinstein-1983} who also initiated  the development of Poisson 
geometry (see also  \cite{Vaisman-1994}).   The concept of  a Poisson structure has  subsequently found numerous applications beyond 
the original focus that was on classical mechanics and differential geometry. 
Poisson structures now appear in a  large variety of contexts starting from string theory,  topological and conformal field theory and integrable systems 
\cite{Babelon-2003,Crainic-2021};  extending to deformation quantization and non-commutative geometry; and all the way to algebraic geometry, representation theory and abstract algebra  \cite{Poisson-2013}.

In this paper we show that  a Poisson structure and in particular the SU(2) Lie-Poisson bracket can also be relevant to the development of effective theory 
descriptions of discrete stringlike objects.  Discrete piecewise linear strings embedded in $\mathbb R^3$ have already appeared 
in models of proteins,  in terms of the C$\alpha$ backbone \cite{Molkenthin-2011}.  They have also important
applications to robotics and 3D virtual reality \cite{hansonbook}. Additional  applications, with more elaborate ambient manifolds,  include the study of
segmented string evolution in de Sitter and anti-de Sitter spaces \cite{Vegh-2021}; see also \cite{arxi-1} and \cite{arxi-2}.

The  paper is arranged as follows. 
Initially, the descendants of the SU(2) Lie-Poisson structure  that relates to the structure of a discrete
piecewise linear polygonal string are considered.  In addition,   the model space and its reduction  in the case of 
the standard SU(2) Lie-Poisson bracket is reviewed.  Then the formalism of the discrete Frenet frames \cite{Hu-2011} 
and  its self-similar hierarchical structure is presented.  Finally, following the results of \cite{Ioannidou-2014}, the
 self-similar structure is converted into a spinor representation, while  the Poisson brackets in terms of 
the SU(2) Lie-Poisson structure are introduced. That way, an infinite hierarchy of Poisson structures
 can be assigned to piecewise linear string as descendants of the canonical SU(2) Lie-Poisson structure.
To conclude, an explicit construction of the first level descendant in this hierarchy is presented in detail.

\section{The model space and the Lie-Poisson structure}
\label{IW}

This preparatory section summarises known results on the model space of SU(2) representations 
and the SU(2) Lie-Poisson structure. 
The  starting point is a four dimensional phase space $\mathbb R^4$ equipped   with a canonical symplectic 
structure and Darboux coordinates ($q_1,p_1,q_2,p_2$) 
\[
\{ p^\alpha, q^\beta \} = - \delta^{\alpha \beta},
\]
 combined  into two complex ones
\begin{equation}
w^\alpha = \frac{1}{\sqrt{2}} (p^\alpha+ i q^\alpha),  \ \ \ \ \ \ \ \ \ (\alpha=1,2).
\la{z12}
\end{equation}

Their  norm is set to be $\rho$, ie.
\begin{equation}
 ||w^1||^2 + ||w^2||^2 = 2\rho,
 \la{rho}
 \end{equation}
while the    associated  Poisson brackets have the simple form,
\begin{equation}
\{ w^\alpha , \bar w^\beta \}  = i \delta^{\alpha\beta}, \ \ \ \ \ \ \ \ \ \  \ \ \ \  \{ w^\alpha , w^\beta \}  
= \{ \bar w^\alpha , \bar w^\beta \} \ = \ 0.
\la{w-bra}
\end{equation}

Next define the three component unit length vector
\begin{equation} 
  t^a \ = \ - \frac{1}{2\rho}
\left( \bar w^1 \  \bar w^2 \right) \sigma^a \left( \begin{matrix} w^1 \\ w^2 \end{matrix} \right),  \ \ \ \ \ \ (a=1,2,3),
\la{n}
\end{equation}
where $\sigma^a$ are the Pauli matrices. 
Then, the $t^a$  components obey the SU(2) Lie-Poisson bracket
\begin{equation}
\{ t^a , t^b \} =  \frac{1}{\rho} \, \epsilon^{abc} t^c,
\la{n-bra}
\end{equation}
associated with the identity
\begin{equation}
\{  t^a , \rho \} = 0.
\la{n-r}
\end{equation}
Therefore, $\rho$ is a Casimir element while  the  phase space  (\ref{z12}) is a model space of SU(2) representations. Note that,   different values of $\rho$ correspond to different representations. The bracket (\ref{n-bra}) determines a Poisson 
structure since:

\vskip 0.2cm

 It is antisymmetric, ie., any two functions $A$ and $B$ satisfy
\begin{equation}
\{ A , B \} = - \{ B , A \}.
\la{PS-1}
\end{equation}

 It  obeys both the Jacobi identity
\begin{equation}
\{ A , \{ B , C\} \} +   \{ B , \{ C, A \} \} + \{ C, \{ A, B\}\} =0
\la{PS-2}
\end{equation}
and the Leibnitz rule
\begin{equation}
\{ A ,  B  C \} \ = \    \{ A , B \}  C + B \{ A, C \}.
\la{PS-3}
\end{equation}
Note that the Jacobi identity coincides with the Schouten bracket of  the Poisson bi-vector field
\begin{equation}
 \Lambda \ = \ \epsilon^{abc} t^c \partial_{a} \wedge \partial_{\hskip 0.2mm b},
\la{Lamb}
\end{equation}
from which the Leibnitz rule  follows directly.  \\

Since the rank of the antisymmetric matrix $\epsilon^{abc} t^c$  is two, 
the bracket in (\ref{n-bra})  does not determine a symplectic structure.
However, the Poisson bracket (\ref{w-bra}) is symplectic with the closed and non-degenerate two-form
\begin{equation}
\omega \ = \ dp_1\wedge dq_1 + dp_2 \wedge dq_2 \ = \  i d w_1 \wedge d w_1^\star  +i  d w_2 \wedge d w_2^\star.
\la{omega} 
\end{equation}

Therefore, a Darboux  coordinate representation of  (\ref{n-bra}) can be derived by introducing the harmonic coordinates
\begin{equation}
\left( \begin{matrix}  w^1 \\ w^2 \end{matrix} \right) \ = \ \sqrt{2\rho}  \left( \begin{matrix}
\vspace{0.1cm} \cos\frac{\theta}{2} \, 
e^{ i (\varphi+ \phi)/2 }  \\ 
\sin\frac{\theta}{2} \, e^{ i (\varphi - \phi)/2 }  \end{matrix} \right), 
\la{zomega}
\end{equation}
and thus, the unit length vector (\ref {n}) simplifies to
\begin{equation}
\mathbf t = \left( \begin{matrix} t^1 \\ t^2 \\ t^3 \end{matrix} \right) \ = \   \left( \begin{matrix} \cos\phi \sin\theta  \\ \sin\phi \sin\theta \\ \cos\theta \end{matrix} \right) .
\la{nproj}
\end{equation}

These coordinates foliate $\mathbb R^4 \sim \mathbb R^1 \times \mathbb S^3 \sim \mathbb R^1 \times \mathbb S^1 \times \mathbb S^2$ where 
 ($\varphi, \phi, \theta $) are the angular coordinates and $\sqrt{2\rho}$ the radii.
That way, the symplectic two-form (\ref{omega}) becomes 
\begin{equation}
\omega 
\ = \   d\rho \wedge d\varphi + \cos\theta d \rho \wedge d\phi + \rho\,   d\cos\theta \wedge d\phi \ \equiv \   d\rho \wedge d\varphi + d (\rho \cos\theta) \wedge d\phi,
\la{omega-2d}
\end{equation}
with the only  non-vanishing Poisson brackets given by
\begin{equation}
 \{ \rho, \varphi   \}  \  = \  -1, \ \ \ \ \ \ \ \ \ \ \ \ \  \{ \rho \cos\theta , \phi  \}  \ = \ -1.
 \la{r4-bra}
\end{equation}

Finally, by setting 
\begin{equation}
\chi  \ =  \  \epsilon  \varphi, 
\la{wi-1}
\end{equation}
and taking  the  In\"on\"u-Wigner contraction  limit ($\epsilon \to 0$) of the system (\ref{r4-bra}), only the second bracket  survives.
The latter  corresponds to the symplectic  Poisson bracket on $\mathbb S^2$ together with its 
closed two-form (unique up to coordinate changes),  that coincides with the last term in (\ref{omega-2d}).
Note that  the coordinate $\rho$ appears only  as a  Casimir element of the Lie-Poisson bracket. Thus, for simplicity, in what follows $\rho=1$.

%
%
%
%
\section{Discrete Frenet Equation and  self-similarity}

\subsection{Vector representation of the discrete Frenet frames}

In this section  descendants of the SU(2) Lie-Poisson bracket defined by  (\ref{n-bra}), that  arise  in connection of open and   piecewise linear polygonal strings ${\Vx}(s) \in \mathbb R^3$, are constructed. 
To set the stage, let $s$ be  the arc length parameter with values $s \in [0,L]$ while  $L$ is the  length of the string.  
Also,  ${\mathscr V}_i$ with $i = 0,...,n$ are the  vertices   that characterise the string  located at the points
${\Vx}(s_i) = {\Vx}_i$. 
Then, neighbouring vertices are connected by the  line segments
\[
{\Vx}(s) \ = \ \frac{ s-s_{i} } {s_{i+1} - s_i} \, {\Vx}_{i+1}  \ - \ \frac{ s - s_{i+1}}  {s_{i+1} - s_i} \, {\Vx}_{i} ,\ \ \ 
\ \ \ \ \ \ s\in(s_{i}, s_{i+1}),
\]
and  are separated by the distances
\[
|\Vx_{i+1} - \Vx_i | = s_{i+1} - s_i \ \equiv
\ \Delta_i.
\]

The discrete Frenet frames are defined by the orthogonal triplets $(\mathbf t,\mathbf n,\mathbf b)_i$ 
 at the vertices ${\mathscr V}_{i}$ as follows: The unit length tangent vectors  $\mathbf t_i $ 
point from  ${\mathscr V}_{i}$ to  ${\mathscr V}_{i+1}$ 
\begin{equation}
\mathbf t_i \ = \ \frac{1}{\Delta_i}( \Vx_{i+1}-\Vx_i),
\la{defti} 
\end{equation}
 the  unit length binormal vectors are
\begin{equation}
\mathbf b_i=\fr{\mathbf t_{i-1}\times \mathbf t_i}{|\mathbf t_{i-1}\times \mathbf t_i |},
\la{b-vec}
\end{equation}
and the unit length normal vectors $\mathbf n_i$ are computed from
\begin{equation}
\mathbf n_i \ = \ \mathbf b_i \times \mathbf t_i \ = \ \fr{-\mathbf t_{i-1}+\left(\mathbf t_{i-1}\cdot \mathbf t_{i}\right)\mathbf t_i\ }{|\mathbf t_{i-1}+\left(\mathbf t_{i-1} \cdot \mathbf t_{i}\right)\mathbf t_i|}.
\la{n-vec}
\end{equation}
In addition,  the   transfer matrix ${\cal R}_{i+1,i}$ 
 maps the discrete Frenet frames between the neighbouring  vertices ${\mathscr V}_{i}$ and ${\mathscr V}_{i+1}$
\begin{equation}
\left(\begin{matrix} \mathbf n\\ \mathbf b \\ \mathbf t \end{matrix} \right)_{i+1}
\ = \  {\mathcal R}_{i+1,i} \left(\begin{matrix} \mathbf n\\ \mathbf b \\ \mathbf t \end{matrix} \right)_i
\ = \ 
\left(\begin{matrix}  \cos \tau  \cos \kappa & \sin \tau \cos \kappa  &-\sin \kappa \\
-\sin \tau & \cos \tau &0\\
\cos\tau \sin \kappa  & \sin \tau \sin \kappa &\cos \kappa \
\end{matrix} \right)_{i}
\left(\begin{matrix} \mathbf n\\ \mathbf b \\ \mathbf t \end{matrix} \right)_i.
\la{DF}
\end{equation}
Here  $\kappa_{i+1}$ is the bond angle and $\tau_{i+1}$ is  the 
torsion angle. 
[Note that,   the transfer matrix ${\mathcal R}_{i+1,i} \in$SO(3) engages 
only two of the Euler angles ($\kappa,\tau$)$_i$ since the 
third Euler angle becomes removed by  the orthogonality of $\mathbf b_i$ and $\mathbf t_{i-1}$.]

The torsion and bond angles ($\kappa_i,\tau_i$) are expressible in terms of the tangent vectors only. This observation follows directly from  equation (\ref{DF}) since
\begin{equation}
\cos \kappa_i \ = \ \mathbf t_{i+1}\cdot \mathbf t_i,
\la{kappa-bra-1}
\end{equation}
while
\begin{equation}
\cos \tau_i \ = \ \mathbf b_{i+1}\cdot \mathbf b_i \ = \ 
\fr{\mathbf t_{i}\times \mathbf t_{i+1}}{|\mathbf t_{i}\times \mathbf t_{i+1} |} \cdot
\fr{\mathbf t_{i-1}\times \mathbf t_i}{|\mathbf t_{i-1}\times \mathbf t_i |}.
\la{tau-bra-1}
\end{equation}
In addition, the bond angle engages three vertices while the torsion
angle engages four vertices along the string. 

The aforementioned construction can be extended into an infinite hierarchy (for an infinite length string)  in a self-similar manner. 
To do so the transfer matrix (\ref{DF}) is used to introduce a $2^{nd}$ level 
orthonormal  triplet of vectors ($\mathbf T, \mathbf N, \mathbf B$)$_i$.
The  components of the vector $\mathbf T_i$ are defined in terms of the last row of (\ref{DF})
\begin{equation} 
\mathbf T_i \ = \ \left( \begin{matrix} \cos\tau_i \sin \kappa_i  \\ \sin \tau_i \sin \kappa_i \\ \cos \kappa_i
\end{matrix} \right),  
\la{2ndT}
\end{equation}
while the corresponding $2^{nd}$ level  binormal and normal vectors, in analogy with (\ref{b-vec}) and (\ref{n-vec}), are  defined as
\begin{equation}
\mathbf B_i \ = \ \fr{\mathbf T_{i-1}\times \mathbf T_i}{|\mathbf T_{i-1}\times \mathbf T_i |}, \ \ \ \ \ \ \ \ \ \ \ \ \ 
\mathbf N_i \ = \ \fr{-\mathbf T_{i-1}+\left(\mathbf T_{i-1}\cdot \mathbf T_{i}\right)\mathbf T_i\ }{|\mathbf T_{i-1}+\left(\mathbf T_{i-1} \cdot \mathbf T_{i}\right)\mathbf T_i|}.
\la{BN-vec}
\end{equation}
Then the corresponding equation  (\ref{DF}) determines  the $2^{nd}$-level transfer matrix
\begin{equation}
\left(\begin{matrix} \mathbf N\\ \mathbf B \\ \mathbf T \end{matrix} \right)_{i+1}
\ = \  {\mathcal R}_{i+1,i} \left(\begin{matrix} \mathbf N \\ \mathbf B \\ \mathbf T \end{matrix} \right)_i
\ \equiv 
\ 
\left(\begin{matrix}  \cos \mathcal T  \cos \mathcal K & \sin \mathcal T  \cos \mathcal K  &-\sin  \mathcal K \\
-\sin \mathcal T & \cos \mathcal T &0\\
\cos\mathcal T\sin  \mathcal K & \sin \mathcal T \sin \mathcal K &\cos \mathcal K \
\end{matrix} \right)_{i}
\left(\begin{matrix} \mathbf N\\ \mathbf B \\ \mathbf T \end{matrix} \right)_i.
\la{DF2}
\end{equation}
with ($\mathcal K, \mathcal T$)$_i$ the $2^{nd}$-level bond and torsion angles  evaluated in terms of the  $2^{nd}$-level $\mathbf T_i$  in analogy to equations (\ref{kappa-bra-1}) and (\ref{tau-bra-1}).

The construction can be extended to the next level. That is, using the last row of (\ref{DF2}) the formulation  (\ref{2ndT})  is 
used  to  introduce  the $3^{rd}$-level tangent vectors. From these,  
the $3^{rd}$ level  vectors  (\ref{BN-vec}) and   transfer matrix (\ref{DF2}) are obtained.
The construction can then be  continued to higher levels (in a self-similar manner) and thus,  an infinite hierarchy is obtained. 
In particular, every vector and angle that appears in this self-similar hierarchy, can be expressed recursively  
in terms of the initial tangent vectors  $\mathbf t_i$.

\subsection{Spinor representation of the discrete Frenet equation}

In this section the spinorial form of  the discrete Frenet equation (\ref{DF})  is presented. 
To do so, a two component spinor is assigned to each link that connects the vertices  ${\mathscr V}_i$ and   ${\mathscr V}_{i+1}$, that is,
\begin{equation}
\psi_{i} =  \left( \begin{matrix} \vspace{0.1cm} z_{1}  \\ z_{2} \end{matrix}\right)^i.
\la{psi1}
\end{equation}
The $z_{\alpha}^{i}$ (for $\alpha=1,2$) are  complex variables assigned to the link. 
Then, the unit length tangent vectors $\mathbf t_i$ can be expressed in terms of the spinors from  a relation akin that in (\ref{n})
\begin{equation}
\psi_i^\dagger  \hat \sigma \psi_i \ = \ \sqrt{g_i}\, \mathbf t_i,
\la{t-rel}
\end{equation}
where $\hat\sigma = (\sigma^1, \sigma^2, \sigma^3)$ are the Pauli matrices, $\mathbf t_i$ is the
discrete tangent vector (\ref{defti}) and $\sqrt{g_i}$ is the  scale factor,
\begin{equation}
\sqrt{g_i} \equiv \left( |z_{1}|^2 + |z_{2}|^2\right)^i.
\la{sqrtg}
\end{equation}

The difference to equation (\ref{rho}) should be noted.  From the definition (\ref{t-rel}) and using (\ref{psi1})
one can easily derive that
\begin{eqnarray}
z_1^i \ =\   \sqrt{ \fr{g_i} {2} }  
 \left[\sqrt{t_1-it_2}\left(\fr{1+t_3}{1-t_3}\right)^{1/4}\right]^{i} ,
\nonumber \\
z_2^i\ = \ 
\sqrt{ \fr{g_i} {2} }  
 \left[\sqrt{t_1+it_2}\left(\fr{1-t_3}{1+t_3}\right)^{1/4}\right]^{i},
\la{zt}
\end{eqnarray}
while  in terms of  the local coordinates (\ref{nproj}) one obtains 
\begin{equation}
 \left( \begin{matrix} z_1 \\ z_2 \end{matrix} \right)^{i} \ = \ 
\sqrt{g_i} \, 
 \left( \begin{matrix}
\vspace{0.1cm} \cos\frac{\theta}{2} \, 
e^{ i \phi/2 }  \\ 
\sin\frac{\theta}{2} \, e^{ -i \phi/2 }  \end{matrix} \right)^{i}.  
\la{z1z2}
\end{equation}
In analogy to  (\ref{zomega})  the  value of the overall factor $\sqrt{g_i}$ can be changed and 
let us  (for simplicity) set $g_i=1$.

Next the  conjugation operation ${\mathscr C}\,$ is introduced to create the conjugate spinor $\bar\psi_i$, 
\begin{equation}
{\mathscr C} \,  \psi_i \ = \  -i \sigma_2 \psi_i^\star \ = \  \left( \begin{matrix} - \bar z_2 
\\ \  \ \bar z_1 \end{matrix}\right)^i \
 \equiv  \  \bar \psi_i ,
\la{psi2}
\end{equation}
so that 
\[
\psi_i ^\dagger  \bar\psi_i \ = \ 0.
\]

Together the two spinors $\psi_i$ and $\bar\psi_i$  define the   2$\times$2 matrix 
\begin{equation}
{\mathfrak u}_i = \left( \begin{matrix} z_1 & -\bar z_2 \\ z_2 & \ \  \bar  z_1 \end{matrix} \right)^i,
\la{u}
\end{equation}
where
\[
\psi _i  =   {\mathfrak u}_i  
\left( \begin{matrix} 1 \\ 0 \end{matrix} \right), \ \ \ \ \\\ \ \ \ \  \bar \psi_i  =   {\mathfrak u}_i  
\left( \begin{matrix} 0 \\ 1 \end{matrix} \right).
\]

Finally, to derive the spinorial discrete Fernet equation in a matrix form, 
a Majorana spinor is constructed from the two spinors  (\ref{psi1}) and (\ref{psi2}) by setting 
\begin{equation*}
\Psi_i = \left( \begin{matrix}  -\bar\psi\\  \ \ \psi\end{matrix} \right)^i,
\label{majspi}
\end{equation*}
one can now introduce a  spinorial transfer matrix $\mathcal U_{i+1,i}$ that relates the Majorana spinors at  the
neighbouring links as
\begin{equation}
\Psi_{i+1} \ = \  \mathcal U_{i+1}^\dagger \ \Psi_i.
\la{discfre1}
\end{equation}
Equation (\ref{discfre1}) is the so-called spinorial discrete Frenet equation; 
In analogy to (\ref{u})  the matrix $\mathcal U_{i+1,i}$ can be expressed in terms 
of the vertex  variables $Z^{i}_{a}$ (for $a=1,2$): 
\begin{equation}
\mathcal U_{i} \ = \ \left( \begin{matrix}  Z_{1} 
& - \bar Z_{2}\\   Z_{2}& \ \ \bar Z_{1}\end{matrix} \right)^i.
\la{matU}
\end{equation} 

The  link $(z_1,z_2)^i$ and the vertex $(Z_1,Z_2)^i$  variables  are connected through the  
discrete Frenet equation (\ref{discfre1}). In particular, 
\be
Z_1^{i+1}&=&\bar{z}_1^i\,z_1^{i+1}+\bar{z}_2^i\,z_2^{i+1}\nonumber\\
Z_2^{i+1}&=&z_1^i\,z_2^{i+1}-z_2^i\,z_1^{i+1}.
\la{Z}
\ee
and the choice $\sqrt{g_i}=1$ in (\ref{sqrtg}) gives $(|Z_1|^2+|Z_2|^2)^i=1$. 

In analogy with (\ref{DF2}), one can  introduce  a $2^{nd}$ level spinor variables, with the ensuing $2^{nd}$ level spinorial Frenet equation. The  construction can be repeated to higher levels, in a self-similar manner,  to obtain an infinite hierarchy of spinorial discrete Frenet equations. Notably, all quantities that appear in this hierarchy can be written in terms of the complex variables (\ref{zt}), recursively.

\section{Descendants of the SU(2) Lie-Poisson bracket }

In the case of the discrete Frenet frames,  the entire self-similar hierarchy can be  constructed 
recursively in terms of the initial tangent vectors (\ref{defti}).  
As a consequence, one can also introduce Poisson structures at all levels of the hierarchy;
recall that the SU(2) Lie-Poisson brackets (\ref{n-bra})  imposed on the tangent vectors (\ref{defti}) 
take the simple form
\begin{equation}
\{ t^a_i , t^b_j \} = \frac{1}{\Delta_i} \delta_{ij} \epsilon^{abc} t^c_i,
\la{t-bra}
\end{equation}
where  $\Delta_i$ are identified as Casimir elements and for  convenience the value $\Delta_i=1$ is chosen.

Equivalently,  the spinor realisation of the hierarchy 
can be expressed recursively in terms of the complex link variables (\ref{psi1}).
Indeed,  from (\ref{t-bra}) it is straightforward to show that the link variables (\ref{zt}) satisfy the following 
algebra
\begin{eqnarray}
\vspace{0.1cm} \nonumber
\{z_\alpha^i, \bar{z}_\alpha^j\} & = & \fr{i}{4}\ \delta_{ij}, \ \ \ \  \alpha =1,2, \\  
\nonumber
\{z_1^i , z^j_2\} &=& -\fr{i}{8}\left(\fr{|z_1|^2-|z_2|^2}{\sqrt{|z_1|^2  |z_2|^2}}\right)^{i} \delta_{ij},
\la{z-bracs} \\
\{z_1^i,\bar{z}_j^2\} &=& -\fr{i}{8} \fr{1}{\left(\bar{z}_1z_2\right)^i} \ \delta_{ij} .
\end{eqnarray}

While it is clear that the Poisson brackets of all the quantities that appear in the self-similar 
hierarchy can be evaluated recursively in terms of  (\ref{t-bra})
it is not obvious that  the Poisson brackets of all the components of $\mathbf T_i$  that appear at a  
given higher level of the hierarchy, form a closed algebra. 
If  this is  the case,  a method  is obtained to systematically generate new Poisson structures,
as higher level descendants of the original  SU(2) Lie-Poisson structure.  
In what follows,  starting from  the spinor representation (\ref{z-bracs}) of the SU(2) Lie-Poisson bracket it is demonstrated by an explicit computation that this is the case.
To do so,   the Poisson brackets of the vertex variables (\ref{Z}) are evaluated.
In particular, they are employed as coordinates  to  define a Poisson structure in terms of the pertinent  
Poisson bi-vector, that is, 
\begin{equation}
\Lambda(Z,\bar Z) \ = \ \Omega^{\mu\nu}(Z^\alpha_i,\bar Z^\alpha_i) \partial_\mu \wedge \partial_\nu \ \ \ \ \ \ \ \mu,\nu \sim (\alpha,i).
\la{Sch}
\end{equation}
After some lengthy algebra it is found that the only non-vanishing brackets of the vertex variables (\ref{Z})  are the following 
\be
&&\{Z_1^{i+1},Z_1^i \}=\fr{i}{2}Z_2^{i+1}\bar{Z}_2^i-\fr{i}{8}\Lambda^i\left(Z_1^{i+1}\bar{Z}_2^i-Z_2^{i+1}Z_1^i\right),\nonumber\acc
&&\{Z_1^{i+1},Z_2^i \}=-\fr{i}{2}Z_2^{i+1}\bar{Z}_1^i+\fr{i}{8}\Lambda^i \left(Z_1^{i+1}\bar{Z}_1^i+Z_2^{i+1}Z_2^i\right),\nonumber\acc
&&\{Z_1^{i+1},\bar{Z}_1^i \}=-\fr{i}{8}\Lambda^i \left(Z_1^{i+1}Z_2^i+Z_2^{i+1}\bar{Z}_1^i\right), \nonumber\acc
&&\{Z_1^{i+1},\bar{Z}_2^i \}=-\{Z_2^{i+1},Z_1^{i}\},\nonumber\acc
&&\{Z_2^{i+1},Z_1^{i} \}=-\fr{i}{8}\Lambda^i\left(Z_1^{i+1}Z_1^i-Z_2^{i+1}\bar{Z}_2^i\right),\nonumber\acc
&&\{Z_2^{i+1},Z_2^i \}=\{Z_1^{i+1},\bar{Z}_1^i \},\nonumber\acc
&&\{Z_2^{i+1},\bar{Z}_1^i \}=\fr{i}{2}Z_1^{i+1}Z_2^i+\fr{i}{8}\Lambda^i\left(Z_1^{i+1}\bar{Z}_1^i+Z_2^{i+1}Z_2^i\right),\nonumber\acc
&&\{Z_2^{i+1},\bar{Z}_2^i \}=-\fr{i}{2}Z_1^{i+1}Z_1^i+\fr{i}{8}\Lambda^i\left(Z_1^{i+1}\bar{Z}_2^i-Z_2^{i+1}Z_1^i\right),\ee
\be&&\{Z_1,\bar{Z}_1 \}^{i+1}=\fr{i}{8}\,\Lambda^i\,\left(Z_1\bar{Z}_2+\bar{Z}_1Z_2\right)^{i+1}+\fr{i}{8}\,\Lambda^{i+1}\,\left(Z_1 Z_2+\bar{Z}_1\bar{Z}_2\right)^{i+1},\nonumber\acc
&&\{Z_1,Z_2\}^{i+1}=\fr{i}{2}\left(Z_1Z_2\right)^{i+1}-\fr{i}{8}\Lambda^{i+1}-\fr{i}{8} \Lambda^{i} \left(Z_1^2-Z_2^2\right)^{i+1},\nonumber\acc
&&\{Z_1,\bar{Z}_2\}^{i+1}=-\fr{i}{2}\left(Z_1\bar{Z}_2\right)^{i+1}-\fr{i}{8}\Lambda^{i}-\fr{i}{8} \Lambda^{i+1} \left(Z_1^2-\bar{Z}_2^2\right)^{i+1},\nonumber\acc
&&\{Z_2,\bar{Z}_2 \}^{i+1}=i\left|Z_1^{i+1}\right|^2+\fr{i}{8}\,\Lambda^i\,\left(Z_1\bar{Z}_2+\bar{Z}_1Z_2\right)^{i+1}-\fr{i}{8}\,\Lambda^{i+1}\,\left(Z_1 Z_2+\bar{Z}_1\bar{Z}_2\right)^{i+1}.
\la{result}
\ee
where the parameter $\Lambda^i$ is real (ie., $\Lambda^i=\bar{\Lambda}^i$) and is defined by the dual form  in terms of the vertex variables either at the $i^{th}$ or at the $i+1^{th}$ vertex$^{[1]}$\footnotetext[1]{This is proven in  the Appendix A; due to  (\ref{L2})}. That is, 
\be\Lambda^i
&=&
\left(\fr{\bar{Z_1}^2-Z_1^2+\bar{Z_2}^2-Z_2^2}{\bar{Z}_1Z_2-Z_1\bar{Z}_2}\right)^{i}\\
&=&\left(\fr{\bar{Z_1}^2-Z_1^2-\bar{Z_2}^2+Z_2^2}{Z_1Z_2-\bar{Z}_1\bar{Z}_2}\right)^{i+1}.
\ee

Furthermore, one can check that the following identities are satisfied
\begin{eqnarray*}
\{|Z_1|^2+|Z_2|^2,Z_1\}^i \  & = & \  \{|Z_1|^2+|Z_2|^2, \bar Z_1\}^i \ = \ 0, \\
\{|Z_1|^2+|Z_2|^2 , Z_2\}^i \ & = &  \{|Z_1|^2+|Z_2|^2 , \bar Z_2\}^i \ = 0, \\
\{\left(|Z_1|^2+|Z_2|^2\right)^{i+1},Z_1^i\} \ & = & \ \{\left(|Z_1|^2+|Z_2|^2\right)^{i+1},\bar Z_1^i\} \ = \ 0, \\
\{\left(|Z_1|^2+|Z_2|^2\right)^{i+1},Z_2^i\} \ & = & \ \{\left(|Z_1|^2+|Z_2|^2\right)^{i+1},\bar Z_2^i\} \ = \ 0, \\
\{\left(|Z_1|^2+|Z_2|^2\right)^i,Z_1^{i+1}\}\ & = & \ \{\left(|Z_1|^2+|Z_2|^2\right)^i,\bar Z_1^{i+1}\} \ = \ 0, \\
 \{\left(|Z_1|^2+|Z_2|^2\right)^{i},Z_2^{i+1}\} \ & = & \  \{\left(|Z_1|^2+|Z_2|^2\right)^{i}, \bar Z_2^{i+1}\} \ = \ 0.
\end{eqnarray*}
Thus $ |Z^i_1|^2+|Z^i_2|^2 $ are Casimir elements of the derived algebra (\ref{result}). [Note that, this result is expected,
 due to the form of the vertex variables defined in  (\ref{Z})].

To sum up,   The relations (\ref{result}) determine a closed, albeit nonlinear, Poisson bracket algebra that 
obeys the Jacobi identity and
the Leibnitz rule, as can be concluded either by general arguments or by explicit evaluation of the  Schouten bracket of 
the pertinent Poisson bi-vector (\ref{Sch}). 
 In particular, the Poisson brackets   (\ref{result})  determine a Poisson structure  that is a proper 
 descendant of the initial SU(2) Lie-Poisson  structure.  The construction can be extended   to all levels of the 
 hierarchy in a self-similar way as explained above. Therefore,  an infinite hierarchy of Poisson structures 
 as descendants of the SU(2) Lie algebra can be constructed.

\section{Concluding remarks}

In conclusion, it has been shown here that in the case of a piecewise linear polygonal string 
the SU(2) Lie-Poisson structure gives rise to an infinite hierarchy of Poisson structures, as its descendants.
 Each level of Poisson structures engages  an increasingly number of vertices along the string, thus they are different. 
It has been shown by an explicit construction of the first level descendant, 
that the spinor representation of the Lie-Poisson bracket is a 
computationally tractable realisation. 
The novel Poisson structure that has been constructed explicitely, engages a chain of four vertices along the 
string (three links), and the higher level descendants  engage  an increasing number of vertices. 

\section{Acknowledgements}

 TI  is supported by COST Action CA17139.
 AJN is supported by the Carl Trygger Foundation Grant CTS 18:276, by the Swedish Research Council under Contract No. 2018-04411, and by COST Action CA17139. 
 Nordita is supported in part by Nordforsk.
 
 \appendix
\section{Link Vs Vertex Variables}
 Directly from (\ref{Z}) the following systems are also satisfied
\be
\ds z_1^{i+1}=z_1^i\,Z_1^{i+1}-\bar{z}_2^i\, Z_2^{i+1}\acc
z_2^{i+1}=z_2^i\, Z_1^{i+1}+\bar{z}_1^i\, Z_2^{i+1}
\dst \ \ \ \ \ \LRa \ \ \ \ \
\ds z_1^{i}=\left(z_1\,\bar{Z}_1+\bar{z}_2\, Z_2\right)^{i+1}\acc
z_2^{i}=\left( z_2\, \bar{Z}_1-\bar{z}_1\, Z_2\right)^{i+1}.\label{bas}
\dst
\ee
where  $|z_1^i|^2+|z_2^i|^2=1$. Note that, by  definition due to (\ref{zt}) the link variables satisfy the identity $\left(z_1z_2\right)^i\equiv\left(\bar{z}_1\bar{z}_2\right)^i$ which is not true for the vertex variables.

Due to  (\ref{bas}) it is easy to prove that  
\be
\left(\fr{|z_1|^2-|z_2|^2}{z_1z_2}\right)^{i+1}=\left(\fr{\bar{Z_1}^2-Z_1^2+\bar{Z_2}^2-Z_2^2}{\bar{Z}_1Z_2-Z_1\bar{Z}_2}\right)^{i+1}, \ \ \ \ 
\left(\fr{|z_1|^2-|z_2|^2}{z_1z_2}\right)^{i}=\left(\fr{\bar{Z_1}^2-Z_1^2-\bar{Z_2}^2+Z_2^2}{Z_1Z_2-\bar{Z}_1\bar{Z}_2}\right)^{i+1}\label{L2}\ee

\end{document}